
\magnification=\magstep1
\font\bigbfont=cmbx10 scaled\magstep1
\font\bigifont=cmti10 scaled\magstep1
\font\bigrfont=cmr10 scaled\magstep1
\vsize = 23.5 truecm
\hsize = 15.5 truecm
\hoffset = .2truein
\baselineskip = 14 truept
\overfullrule = 0pt
\parskip = 3 truept
\def\frac#1#2{{#1\over#2}}


%
%

\centerline{\bigbfont QUANTUM POINT CONTACTS AND BEYOND: NEW RESULTS}
\centerline{\bigbfont ON MESOSCOPIC CONDUCTANCE AND FLUCTUATIONS}

\vskip 20 truept

\vskip 8 truept
\centerline{
{\bigifont Mukunda P. Das${}^{\dagger}$,
Frederick Green${}^{\dagger\ddagger}$},
{\bigrfont and}
{\bigifont Jagdish S. Thakur${}^{*}$}
}
\vskip 8 truept
\centerline{${\dagger}$\bigrfont Department of Theoretical Physics}
\vskip 2 truept
\centerline{\bigrfont Research School of Physical Sciences and Engineering}
\vskip 2 truept
\centerline{\bigrfont The Australian National University}
\vskip 2 truept
\centerline{\bigrfont Canberra ACT 0200, Australia}
\vskip 8 truept
\centerline{${\ddagger}$\bigrfont
School of Physics, The University of New South Wales}
\centerline{\bigrfont Sydney NSW 2052, Australia}
\vskip 8 truept
\centerline{${*}$\bigrfont Department of Electrical
and Computer Engineering}
\vskip 2 truept
\centerline{\bigrfont Wayne State University}
\vskip 2 truept
\centerline{\bigrfont Detroit, Michigan 48202, USA}

\vskip 12 truept

\vskip 16 truept
\centerline{\bf 1. INTRODUCTION}
\vskip 12 truept

After a century of development, during which the microscopic foundations of
transport kinetics were established as one of the great achievements
of modern physics, the last decade and a half has seen fresh, indeed
revolutionary, progress in the understanding of electrical conduction.
This progress has never been so rapid as in the mesoscopic realm.
It can be said that, in the place of more traditional
microscopic perspectives, a succession of novel, simpler, and more
succinct approaches have arisen to bolster our notions of
mesoscopic transport physics.

The guiding spirit of this new
philosophy is rightly identified as Rolf Landauer. Other
pioneers, such as B\"uttiker, Beenakker, and Imry,
have been foremost in deepening and extending
Landauer's phenomenological insights into their final,
contemporary form [1-5].
The astonishing success of these highly influential mesoscopic
models elicits two questions:

\item{(i)}
what is the relationship between the ``new'' viewpoint
of electron transport and the established, microscopic one?

\item{(ii)}
Are there reasons to continue placing confidence
in the ``old'' microscopic theory as the source of
fresh insights into mesoscopics?

Our review proposes answers to these questions. In the first
instance, an objective assessment of the principles underlying
the physics of mesoscopic transport leads us to
state -- with some assurance -- that
{\it conservative microscopic methods remain the only reliable
basis for solving mesoscopic problems}.
For example, Fermi-liquid theory [6, 7]
readily accounts for some major and puzzling experimental
results in quantum point contacts.
Second, a fully conserving approach lets one predict
new effects to be sought out in new experiments, using
available methods and device structures. These effects are
not addressed by recently developed phenomenologies.

The credibility of any mesoscopic model, new or old, hinges
above all upon its respect for the conservation laws. A mesoscopic
theory that violates these basic statements does not
make physical sense. It hardly matters if such a description
were to claim for itself some special sort of compactness,
simplicity, and intuitive appeal. If it cannot secure conservation,
nothing else can commend it.$^1$
\footnote{}{${^1}$ {One recalls Einstein's sharpened form
of Occam's Razor: ``Everything should be made as simple as
possible, but not simpler.''}}

The centrality of energy, charge, and number conservation [6]
will form the connecting thread of our review.
In Section 2 we briefly recall the recent history of mesoscopics,
setting the stage for our microscopically based critique; this
analysis rests on the close relationship between microscopic
conservation and the {\it open-boundary conditions} dominant in
mesoscopic electron transport. In Sec. 3 we apply that knowledge
to an exemplary case:
the conductance of a one-dimensional quantum point contact (QPC).
We will see how Landauer's celebrated expression for
quantized conductance emerges straight from the standard
Kubo-Greenwood theory [8-10] applied to an open mesoscopic wire,
in which elastic {\it and} dissipative scattering are both vital elements.
The theme is filled out, in Sec. 4, in our kinetic description of
nonequilibrium current fluctuations (noise) of a driven
quantum point contact. Noise and conductance are intimately
linked; each manifests the {\it two-body} electron-hole
dynamics at the heart of metallic conduction [6, 11].
We discuss the resolution of a long-standing experimental
problem involving the excess noise of a QPC [12, 13], which other models
have been unable to address. This has far-reaching implications
for the way in which measurements of fluctuation phenomena are
interpreted to reveal the microscopic details of
low-dimensional electron transport.
In Sec. 5 we foreshadow some novel, testable consequences
of the physics of the preceding Section.
Finally, we gather our thoughts in Sec. 6.




\vskip 16 truept
\centerline{\bf 2. LANDAUER THEORY: A SHORT HISTORY}
\vskip 12 truept

In 1957, Rolf Landauer published a different and -- to some
of the leading transport gurus of the epoch -- subversive
interpretation of metallic resistivity [14].
Landauer envisioned the current, rather then the electromotive
voltage, as the stimulus by which resistance is manifested [15].
The measured voltage is simply the macroscopic
effect of the carriers' inevitable encounters with the localized
scattering centers within a conductor. Around any such scatterer
the carrier flux resembles a phenomenological ``diffusive'' flow,
set up by the density difference between the upstream and downstream
electron populations. In this purely passive scenario,
energy dissipation does not enter.

The Landauer theory describes electron transport in an environment
of scatterers that are {\it purely elastic}. As such
it is not able to address the dynamical mechanisms of energy
dissipation that characterize transport. Yet the Landauer theory,
like any other description of conductance, must satisfy the
fluctuation-dissipation (Johnson-Nyquist) theorem at some level.
If it did not do so, the theorem must be force-fit -- by hand -- to
the model.
The fluctuation-dissipation theorem implies that dissipation
is always present whenever there is resistance.
It is an inescapable element in every theory
of conductance, be it microscopic or phenomenological.
Landauer's conceptual model assigns no role at all to inelastic
dissipation in its handling of scattering-mediated conductance.

The Landauer formula for the conductance $G$ of a quantum point
contact is

$$
G = {2e^2\over h} {\cal T}.
\eqno(1)
$$

\noindent
This version of the formula prescribes the outcome of
a ``two-point'' measurement; that is, the bulk source and drain leads
(needed to supply current to the QPC) are the points between
which the voltage as well as the current is measured.
For an ideal ballistic conductor the transmission
coefficient ${\cal T}$ is unity. In the presence of coherent
backscattering off the contact, or of inelastic
scattering -- or both -- the conductance is
nonideal since ${\cal T} < 1$. Note,
however, that the Landauer model provides no guidance for
computing ${\cal T}$ in the case of coherent elastic
scattering. Moreover the obvious possibility of inelastic
scattering remains permanently out of scope, by construction.

At this point we slightly anticipate what is to come.
Eq. (1) conceals a tantalizing conundrum: how, exactly,
can a driven system shed its excess energy in the Landauer picture?
A mesoscopic channel with conductance given by Eq. (1)
should dissipate the electrical power supplied to it at the rate

$$
P = IV = {2e^2\over h} {\cal T} V^2.
\eqno(2)
$$

\noindent
Yet, to drive this dissipation, there is no dynamical mechanism
that can be identified within the object ${\cal T}$,
central to the accepted picture. How can this be?
The process of transmission is supposed to be purely coherent (elastic).
Elastic collisions conserve energy; they are nondissipative.

This means that Eq. (2) can never be more than an added-on
hypothesis in the Landauer view of transport; one that is devoid of
supporting reason beyond the obviously heuristic observation
that finite conductances dissipate finite electrical energy.
Therefore, because of the complete absence of a {\it microscopic}
nexus between the Landauer formula and
the fluctuation-dissipation relation, Eq. (2),
there is an enormous conceptual gap that needs healing.
Such a gap cannot be closed within the scheme's internal logic.
To resolve the dilemma we have no option but to go back to the
pre-Landauer understanding of quantum transport.

For all the succinctness and empirical success
of Landauer's conception, by his own recounting [16]
the theory languished until its rediscovery and reinterpretation
by a fresh, bold generation of transport physicists.
The main early objection seemed to be
his emphasis on the localized action of the scattering
impurities in resisting current flow, at a time when the theoretical
dogma held that local effects could never be individually probed;
all that one could (and should) do was to compute a spatial
average over ensembles of samples, with a spread of scattering-center
distributions [17].

All this changed radically in the 'eighties, with the
advent of truly mesoscopic sample fabrication. It was now possible to
study not bulk, and hence coarsely grained assemblies, but
individual samples with individual spatial arrangements of scatterers.
Moreover, the phase coherence of the carriers could now be
preserved over the much shorter lengths of the samples,
bringing to the fore the effects of quantum transmission.

Landauer's criticism of bulk averaging was entirely justified;
that particular approximation is no longer meaningful for
truly mesoscopic structures. However, the breakdown
of bulk averaging is, as we have said,
at a very coarse-grained level. It does not impact upon
the fundamental structure of statistical mechanics with its
{\it microcanonical} averaging in configuration space,
over identical runs for the system's response
(or, alternatively, over a set of identical copies
of the device: ergodicity).
The whole of quantum kinetics is built on the microscopic
ensemble procedure, which is universally applicable
at all length scales and dimensionalities [18].

The applicability of statistical mechanics and kinetics to
mesoscopic transport is clearly not subject to Landauer's
restricted argument against bulk spatial averaging.
His reconsideration of its limitations carries no
physical reason to bypass -- let alone supplant -- traditional
microscopic methods in favor of more conjectural treatments.
One such traditional method, which we will apply below in a
new experimental context, is the Landau-Silin quantum-transport
equation for metallic electrons,

Despite the circumscribed nature of Landauer's critique,
there seems to be an informal but widespread perception
that his quasi-diffusive picture (dressed in a single-particle
interpretation of coherent transmission)
supersedes, as it were, the collective efforts of kinetic
theorists through the preceding century of research.$^2$
\footnote{}{$^2$ {Such an impression was never put about by Landauer
who, in the abstract for the reprint of his seminal 1957 paper in
J. Math. Phys. {\bf 37}, 5269 (1996),
interpolated this forthright observation:
`` ...[The] IBM Journal of Research and Development,
is not all that easily located in 1996. As a result
the frequent citations to it often assign content to
that paper which does not agree with reality.''}}
Granted, the old microscopic ways are too labor-intensive
to meet modern demands for a prolific research output.
Further, the folklore has it that microscopics can, at best,
only confirm and enshrine Landauer's far more compact phenomenology.
In any case, it is now believed that mesoscopic transport occurs
exclusively by elastic coherent transmission, to all intents
dissipation-free (even when the fluctuation-dissipation theorem
explicitly says otherwise). It is as if the reality of inelastic
processes beyond that horizon were suddenly inoperative.

The point is not to dwell on the achievements of the past, nor to
speculate on the durability of present phenomenologies
(that is a suitable case for Occam's Razor). Our aim is quite
practical. The {\it minimum} requirement for transport models is
that they be conserving. Physical credibility must be tested,
not by fashion or expediency, but by the same objective criteria
against which theories have always been -- and will always be -- judged.
Let us put the new mesoscopics to the test.

Do quasi-diffusive models genuinely bring a novel understanding to
mesoscopics, an understanding that could not be reached through
the canonical and conservative methods of kinetic theory?
Do quasi-diffusive models genuinely conform to the conservation
laws that apply to open mesoscopic conductors?
The answers to these issues may surprise some readers.

\vskip 16 truept
\centerline{\bf 3. THE LANDAUER FORMULA, WITHOUT PHENOMENOLOGY}
\vskip 12 truept

In the literature there are already numerous
attempts to derive and hence justify Landauer's
phenomenological conductance formula [1, 4].
This is natural, given that the empirical success of the formula
is as compelling as its detailed microscopic underpinning is unclear.
Its true underpinning will be exhibited below.

The best-known derivations of Landauer's result mostly start with
the many-body Kubo-Greenwood conductance formula, which is our
starting point as well. However, those derivations tend to
rely on the same assumptions about boundary conditions
(and the absence of recognizably dissipative processes)
that were made by Landauer himself.
As reputedly microscopic confirmations, they beg the question
of logical circularity and their dependence on the results
that they aim to prove.

The key to mesoscopic transport theory is knowing how
to handle, in a fully conserving way, the physical interactions
between the open device and the macroscopic environment,
mediated by its {\it interfaces}. Before turning to
the microscopic Kubo-Greenwood theory and its straightforward
recovery of Landauer's result, we recall some basic facts
about the boundary conditions for an open electronic conductor.

In mesoscopic transport the role of the macroscopic metallic
leads connected to the sample is paramount.
The leads, or reservoirs, will

\item{(a)} confine all electric fields, be they external or inbuilt,
to a well-defined region comprising the device and its interfaces
with the leads (namely, the connected region where carriers are
appreciably disturbed by the applied current);

\item{(b)} ensure the permanent neutrality of the device
with its interfaces, irrespective of the applied current
(this is the consequence of strong Thomas-Fermi screening
by the conduction electrons in the leads); and

\item{(c)} pin the asymptotic state, to which the driven carriers
in the active volume will tend, to the permanently stable unchanging
(equilibrated and electrically neutral) local distribution
in each reservoir.

\noindent
These conditions, almost self-evident for metallic conduction, are
common to all transport models including those of Landauer type.
They also correspond to the laboratory set-up required in all
measurements of conductance. 

\vskip 12 truept
\centerline{\it a. The Kubo-Greenwood Formula}
\vskip 12 truept

Under the open-boundary conditions above, it was proved
by Sols [10]
that the Kubo-Greenwood conductance formula holds in a form
absolutely identical to the well known case of a
closed electronic conductor with periodicity.
(An independent but physically equivalent proof was given by
Magnus and Schoenmaker [19]). There is a crucial proviso:
the global gauge invariance (charge conservation) of the
open-system Kubo-Greenwood formula is guaranteed
{\it if and only if} an external generator actively
injects and extracts the external current that
energizes the open mesoscopic device.

The physical need for active sources and sinks of current
in the problem is, then, mandated by charge conservation.
It is a fundamental result that should be compared with the very
different requirement of the Landauer model for an open
conductor. There, it is assumed that a phenomenological mismatch
of carrier densities in the leads ``drives'', entirely
passively, the diffusive-like transfer of
electrons from the nominally high-density source lead
to the nominally low-density drain lead (while the quantum
transmission of the individual carriers regulates the net
diffusion rate).

The hypothesis of quasi-diffusive transport stands in clear
contrast with condition (c) above. We note that this already
hints at some internal contradiction, since (c) involves the
asymptotic neutrality of the leads. That {\it transport-independent}
condition is still required for the Landauer model to remain
electrically stable.

The Kubo-Greenwood conductivity is

$$
\sigma(t) = {ne^2\over m^*}\int^t_0 {\cal C}_{vv}(t) dt, 
\eqno(3)
$$

\noindent 
where the velocity-velocity correlation function is 

$$
C_{vv}(t) = {{\langle v(t)v(0) \rangle}\over
             {\langle v(0)^2\rangle}} 
\sim \exp(-t/\tau_m);
\eqno(4)
$$

\noindent
the expectations are taken in the equilibrium state
(giving linear response)
over the full many-body density matrix for the assembly
of active carriers in the channel$^3$,
\footnote{}{$^3$ {The expectations in $C_{vv}$ are
determined by the {\it electron-hole} excitations
in the structure. In its very essence $C_{vv}$ is 
a correlated many-body quantity, impossible to
represent in pure terms of independent single-particle properties.}}
and $\tau_m$ characterizes the dominant decay of the correlations.
The decay parameter includes, on an equal footing, the microscopic
contribution from every physically relevant collision mechanism
[8, 9].
In the long time limit,

$$
\sigma \to {ne^2\tau_m\over m^*}.
\eqno(5)
$$

\noindent
This is the venerable Drude conductivity.

In one dimension, appropriate to a QPC,
the density in terms of Fermi velocity is $n = 2m^*v_{\rm F}/\pi\hbar$.
The conductance over a sample of length $L$ then becomes

$$
G \equiv {\sigma\over L} = {2m^*v_{\rm F}e^2\over \pi\hbar m^*L} \tau_m
= {2e^2\over h}{\left( {2v_{\rm F}\over L} \tau_m \right)}
\equiv {2e^2\over h}{\cal T}_{\rm KG},
\eqno(6)
$$

\noindent
in which the transmission coefficient ${\cal T}_{\rm KG}$
(KG for Kubo-Greenwood) is now proportional
to the ratio of the effective scattering length $v_{\rm F}\tau_m$
to length $L$.

All of the dissipative, many-body scattering effects have been
embedded within $\tau_m$, as well as all the one-body 
coherent potential and elastic impurity scattering.
The interface physics is incorporated into the microscopic
KG conductance formula, as fully and directly
as the physics of the device itself.$^4$
\footnote{}{$^4$
In contrast with the Landauer formula, there is
no need to graft the fluctuation-dissipation
theorem onto this formula as a heuristic appendage;
the open-system conductivity equation (3) is the theorem.
Therefore no ``supplementary hand-waving'' [15] -- no heuristics
of any sort -- is required to make Eqs. (3)--(6)
mesoscopically legitimate. They are valid from within.}

There is nothing in Eq. (6) to forestall its
conformity with the Landauer formula. To be sure,
Eq. (1) is about to appear as a particular case.
Nevertheless, unlike the received derivations of the latter,
the KG formula does not call for hand-waving supplementations
that favor elastic processes over the inelastic ones that are
equally important in real physical situations.
We now demonstrate just how vital inelastic scattering is
for the Landauer formula itself, by constructing
its correct microscopic origins.

\vskip 12 truept
\centerline{\it b. The Landauer Formula Redux}
\vskip 12 truept

\topinsert
\vskip -0.30truecm
\input psfig.sty
\centerline{\hskip0mm\psfig{figure=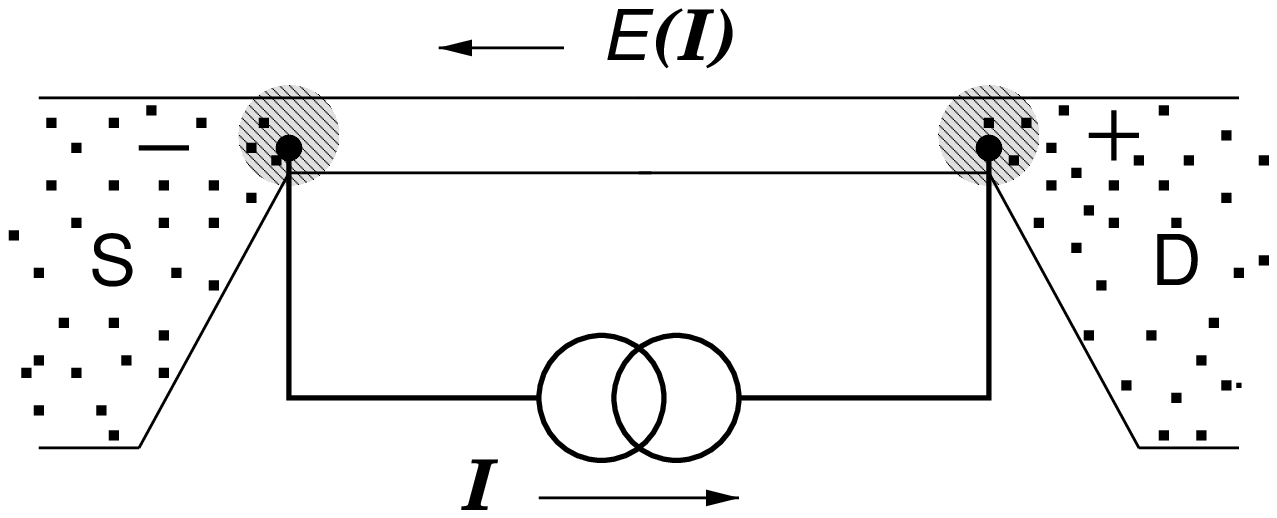,height=4.0truecm}}
%
\vskip 0.25truecm

\noindent
{\bf Figure 1.}
An ideal, uniform ballistic wire. Its diffusive leads
(S, D) are at equilibrium. A paired source and sink of current
$I$ at the boundaries drive the transport. The boundary regions,
separated by distance $L$, are loci for both inelastic and
elastic scattering. Local charge clouds (shaded), induced by
the influx and efflux of $I$, generate the resistivity
dipole potential $V = E(I)L$ between D and S.
\vskip 8truept
\endinsert
\hsize = 15.5truecm

Figure 1 displays the nonequilibrium transport of
current through a conductor, in the present case
a quantum point contact. Note the arrangement whereby
an external generator imposes current flow, while
the outlying macroscopic reservoirs remain permanently
undisturbed at their fixed, equilibrium electron densities
(and their fixed {\it local} chemical potentials).
That is,

\item{$\bullet$}{\it the current through an open conductor
cannot be a function of the asymptotic states in
the leads, and vice versa.}

\noindent
The immunity of the lead states to transport is the
general consequence of global charge conservation
in any open device [10]. It is to be contrasted with the
very different boundary condition invoked by
phenomenological prescriptions: current is
{\it the quasi-diffusive flow of carriers between
reservoir leads when their nominal chemical potentials,
and hence their nominal densities, become mismatched
in the presence of a source-drain voltage}.
Such a statement would clearly contradict
global gauge invariance, since it posits a direct functional
link between the presence of the current and the mismatch in
reservoir populations, set up to sustain it.

The existence of macroscopically large reservoirs outside
the active region sets limits on both the elastic
and inelastic mean free path (MFP). These will be finite
in range even if the channel's mesoscopic core is ballistic
and perfectly collisionless. The presence of defects
in the leads put an upper bound on the elastic MFP.
More to the point, the dynamical effect of the
{\it external} sources and drains for current sets the same
essential bound on the inelastic and elastic MFP.
When nonequilibrium effects, say phonon emission,
set in as the current is increased, the inelastic
MFP will become small compared with the elastic
one, which remains practically unchanged.

We suppose that the ballistic channel has an
operational length $L$. This is identifiable with its
elastic MFP. If the characteristic mean speed of its
carriers is $v_{\rm av}$, then the elastic collision time is

$$
\tau_{\rm el} = {L\over v_{\rm av}}.
\eqno(7a)
$$

\noindent
On the other hand, in a ballistic conductor
the inelastic scattering time cannot exceed $\tau_{\rm el}$
as a limiting value:

$$
\tau_{\rm in} \leq {L\over v_{\rm av}}.
\eqno(7b)
$$

\noindent
The two types of scattering are spatially coextensive,
but temporally independent in their action.
Therefore Matthiessen's rule holds, giving

$$
{1\over \tau_m} = {1\over \tau_{\rm in}} +{1\over \tau_{\rm el}}.
\eqno(8)
$$

\noindent
When the form for $\tau_m$ is substituted in Eq. (6),
the Kubo-Greenwood form for the transmission
factor ${\cal T}_{\rm KG}$ becomes

$$
{\cal T}_{\rm KG} =
{\left( {2v_{\rm F}\over L} \tau_{\rm el} \right)}
{ \tau_{\rm in}\over {\tau_{\rm el} + \tau_{\rm in}} }
\equiv {v_{\rm F}\over v_{\rm av}} {2\zeta\over {1 + \zeta}},
\eqno(9a)
$$

\noindent
where the collision-time ratio is

$$
\zeta = \tau_{\rm in}/\tau_{\rm el} \leq 1.
\eqno(9b)
$$

The result of Eq. (9) is shown in Fig. 2 overleaf. The conductance
of a QPC, in this case consisting of two well separated
one-dimensional subbands, exhibits the classic Landauer
quantization in units of $G_0 = 2e^2/h$
as the density of mobile electrons in the QPC
is swept upwards, and the chemical potential $\mu$ successively
crosses the subband thresholds $\varepsilon_1$ and $\varepsilon_2$.
There are two cases of interest.

\topinsert
\vskip -0.30truecm
\input psfig.sty
\centerline{\hskip15mm\psfig{figure=dgtf2.ps,height=10.0truecm}}
%
\vskip -2.00truecm

\noindent
{\bf Figure 2.}
Conductance of a one-dimensional quantum point contact, computed
with the fully microscopic Kubo-Greenwood model of Eqs. (6)--(9).
We show $G$ scaled to the Landauer quantum $G_0$, as a function of
chemical potential $\mu$ in units of thermal energy [20].
$G$ exhibits strong shoulders as $\mu$ successively
crosses the subband energy thresholds (here at
$\varepsilon_1 = 5k_{\rm B}T$ and $\varepsilon_2 = 17k_{\rm B}T$).
Well above each threshold, the subband electrons are strongly
degenerate and the conductance tends to a well defined
quantized plateau. Much below each threshold, the population
and its contribution to $G$ vanish as
$e^{-(\varepsilon_i-\mu)/k_{\rm B}T}$.
Solid line: $G$ in the ideal ballistic regime,
for which the collision-time ratio $\tau_{\rm in}/\tau_{\rm el}$ is unity.
Chain line: non-ideal case for $\tau_{\rm in}/\tau_{\rm el} = 0.75$.
Note how the increased inelastic scattering brings down the
plateaux. Dotted line: the case of
$\tau_{\rm in}/\tau_{\rm el} = 0.5$. The departure from ideality
is now substantial.
\vskip 8truept
\endinsert
\hsize = 15.5truecm

\item{(a)}{\it Classical limit}.
When the density in one of the subbands $i = 1, 2$ is
very low we have $\mu - \varepsilon_i \ll k_{\rm B}T$ and
$v_{\rm av} \to \sqrt{2m^*k_{\rm B}T}$,
the thermal velocity of Maxwellian particles.
Eq. (5) together with (9a) gives the conductance
contribution in subband $i$

$$
G_i = G_0 {hn\over 4m^*v_{\rm th}} {2\zeta\over 1 + \zeta}
\sim e^{-(\varepsilon_i - \mu)/k_{\rm B}T}.
\eqno(10)
$$

\item{}
\noindent
This contribution vanishes as the subband becomes depleted.

\item{(a)}{\it Degenerate limit}.
When the subband electron density is high, we have
$\mu - \varepsilon_i \gg k_{\rm B}T$
and $v_{\rm av} \to v_{\rm F}$. Then

$$
G_i = G_0 {2\zeta\over 1 + \zeta}.
\eqno(11)
$$

\item{}
\noindent
The kinematic ratio $v_{\rm F}/v_{\rm av}$ goes to
unity, and the conductance reaches a plateau 
exactly as for the phenomenological Landauer formula.

For perfect ballistic transport in a degenerate QPC
(a highly idealized limiting case) one has
$\zeta = v_{\rm F}/v_{\rm av} = {\cal T}_{\rm KG} = 1$.
When $\zeta$ is at its maximum of unity, the elastic
and inelastic scattering times are exactly matched.
Each contributes {\it equally} to the visible transport behavior,
yielding the original Landauer result:

$$
G_i = G_0.
$$

\noindent
Thus his formula is recovered strictly
from the fundamental and much more general KG description,
for the case of an open one-dimensional ballistic channel.

Our microscopic derivation of the Landauer formula, Eq. (1),
makes no appeal to Landauer's phenomenological premises.
Up to now, these assumptions have been regarded as
indispensable to the physics of conductance quantization.
We stress that Eq. (1) emerges naturally and directly
from a fully conserving microscopic treatment. This starts
with the Kubo-Greenwood formula underpinned by
strict boundary-condition requirements, imposed upon the physics
by the gauge invariance of an open electrical conductor.

What now of the conundrum we met in the
previous Section? The conceptual dilemma that has always
confronted the Landauer picture, namely how to conjure
dissipation out of a strictly dissipationless conductance,
simply vanishes away. For, the detailed physical processes
that lead directly to inelasticity and dissipation are fully
included in the microscopic Eq. (6)
side by side with the elastic processes, as a matter of course.

Unlike the Landauer formula, the KG formula tells us how to
compute the conductance in explicit terms of the microscopic
{\it electron-hole pair excitations} of the system.
That is because Eq. (3) embodies the fluctuation-dissipation theorem,
in which the magnitude of the excitations (setting the
loss rate) is modified by the effects of all sources
of scattering on the propagating electron-hole pair.
For an open device, it follows that dissipative and elastic mechanisms
both contribute to the structure of the current fluctuations,
which themselves fix the total collision time ${\tau_m}$,
the transmissivity ${\cal T}_{\rm KG}$, and finally
the conductance $G$ as measured in the laboratory.

Microscopically it makes little sense to rank either one
of the two modes of scattering, inelastic and elastic,
as more ``physical'' than the other in some subjective way.
It makes even less sense to neglect inelastic scattering
altogether in mesoscopics, as is centrally assumed in
diffusive-like approaches. To the contrary we have demonstrated that

\item{$\bullet$} {\it within a strictly conserving framework,
both elastic and inelastic scattering are needed to for
the proper (many-body) description of transport in an open
mesoscopic structure}.

\noindent
Figure 2 further shows that increasing the inelasticity
in the system ($\zeta < 1$)
makes the ballistic QPC nonideal. While preserving the
quantized-step character of $G$ as a function of electron density,
the size of the steps undershoots the ``perfect'' value $G_0$.
Observationally, however, one cannot tell from such plots alone
whether the transmissivity ${\cal T}_{\rm KG}$ really falls
below unity because stronger inelasticity overcomes
the elastic processes, or (conceivably)
because coherent backscattering reduces the elastic relaxation
time $\tau_{\rm el}$ itself while inelasticity plays no role at all.
The phenomenological understanding of nonideality
advances the latter alternative.

From the microscopic standpoint, we have already seen that
there are powerful reasons, based on the conservation laws,
to assert the first possibility over the second.
To uniquely resolve the issue of nonideal transport in a QPC
requires a deeper look at carrier dynamics as revealed by studies of
current fluctuations, and the noise that provides their
experimental signature. We now turn to the microscopic behavior
of fluctuations in a quantum point contact.

\vskip 16 truept
\centerline{\bf 4. QPC NOISE: NEW RESULTS FROM KINETIC ANALYSIS}
\vskip 12 truept

We can easily quantify the open boundary conditions (a)--(c) of
Sec. 3. First: the active region, made up of the driven channel and its
disturbed interfaces with the equilibrated leads (the interfaces
are also the loci of current injection and removal), occupies a
finite volume $\Omega$. {\it Metallic screening} of the fields
internal to $\Omega$ means that it remains independent of the
externally applied current $I$. Second: let the total number of
mobilized electrons in $\Omega$ be $N$. Since global neutrality
holds, the total electronic charge $-eN$ within $\Omega$ is
always {\it compensated} by its positive ionic background,
regardless of how much current is applied. Hence

$$
{d \Omega\over dI} = 0 =
{\left. {{\partial N}\over {\partial I}} \right|}_{\mu}.
\eqno(12)
$$

\vskip 12 truept
\centerline{\it a. Boundary Conditions on Mesoscopic Fluctuations}
\vskip 12 truept

In terms of the local distribution of electrons
$f_{\bf k}({\bf r}, t)$ in wavevector space ${\bf k}$
and real space ${\bf r}$, Eq. (12) immediately
implies that

$$
\int_{\Omega} d^{\nu}r \int {2d^{\nu}k\over (2\pi)^{\nu}}
f_{\bf k}({\bf r}, t) = N =
\int_{\Omega} d^{\nu}r \int {2d^{\nu}k\over (2\pi)^{\nu}}
f^{\rm eq}_{\bf k}({\bf r})
\eqno(13)
$$

\noindent
for all $I$ and all times $t$, and where
$f^{\rm eq}_{\bf k}({\bf r})$ is the equilibrium distribution
within $\Omega$. The dimensionality of the system is $\nu$.
Note that Eqs. (12) and (13) do {\it not} mean that $N$ is
fixed once and for all, and cannot be changed. If, for
example, we alter the thermodynamic conditions so
that the chemical potential changes by $\delta \mu$,
the local electron distribution in the neutral leads will
change its density accordingly to $n(\mu + \delta \mu)$
(the leads' positive background, of course,
must also change to compensate). It follows that
both sides of Eq. (13) are altered by an identical amount:

$$
\int_{\Omega} d^{\nu}r \int {2d^{\nu}k\over (2\pi)^{\nu}}
\delta f_{\bf k}({\bf r}, t) = \delta N =
\int_{\Omega} d^{\nu}r \int {2d^{\nu}k\over (2\pi)^{\nu}}
\delta f^{\rm eq}_{\bf k}({\bf r}).
\eqno(14)
$$

\noindent
This is the {perfect-screening sum rule} [6],
expressed for an open system bounded by its
large metallic leads. Global neutrality
(gauge invariance) guarantees it.

Equation (14) has a striking corollary for the dynamic
fluctuations of an open mesoscopic conductor, and it is this 
result that holds the key to our new interpretation
of quantum-point-contact noise and our new predictions for it.
Recall that the local, mean-square number fluctuation in the
free electron gas is given by

$$
\Delta f^{\rm eq}_{\bf k}({\bf r})
\equiv k_{\rm B}T
{ {\partial f^{\rm eq}_{\bf k}}\over {\partial \mu} }({\bf r})
= f^{\rm eq}_{\bf k}({\bf r})(1 - f^{\rm eq}_{\bf k}({\bf r})).
$$

\noindent
Denoting by $\Delta f_{\bf k}({\bf r}, t)$
the corresponding mean-square number fluctuation in the driven
channel, our key result is that

$$
\int_{\Omega} d^{\nu}r \int {2d^{\nu}k\over (2\pi)^{\nu}}
\Delta f_{\bf k}({\bf r}, t)
= k_{\rm B}T { {\partial N}\over {\partial \mu} }
= \int_{\Omega} d^{\nu}r \int {2d^{\nu}k\over (2\pi)^{\nu}}
\Delta f^{\rm eq}_{\bf k}({\bf r}).
\eqno(15)
$$

\noindent
The consequences of the gauge-invariant boundary conditions
for the fluctuations themselves -- and thus the noise -- now
become clear for a mesoscopic device: Eq. (15) means that

\item{$\bullet$} {\it the total fluctuation strength within
the active region of an open mesoscopic channel is perfectly
independent of the applied current.}

\eject
\vskip 12 truept
\centerline{\it b. Compressibility Sum Rule for Open Systems}
\vskip 12 truept

In the physics of the closed electron gas, the compressibility
sum rule is a well known identity [6]; all viable models of
such a system must satisfy it. We present it here in a new form,
generalized to the open electron gas of a mesoscopic conductor,
such as a QPC to which we will apply it.

Recall the thermodynamic definition of the
electron-gas compressibility $\kappa$:

$$
\kappa
\equiv {\Omega \over N^2}
{\left. { {\partial N}\over {\partial \mu} } \right|}_{\Omega, T}
= {\Omega \over N k_{\rm B}T} {\Delta N\over N},
\eqno(16)
$$

\noindent
where $\Delta N$ is the total mean-square fluctuation
over the region $\Omega$. The compressibility is the inverse of
the average energy density of the system. If it is large, the
system fluctuates readily. If $\kappa$ is small, the system
is stiff and its fluctuations are suppressed.

Empirically, $\kappa$ can be obtained from measurements of the
sound velocity in the electron gas. Eq. (16) relates its
value to the (calculable) magnitude of the
{\it microscopic fluctuations} of the carriers.
Note the close analogy with the fluctuation-dissipation theorem.
There, the empirical conductance $G$ is related to the
magnitude of the current fluctuations in the system.

We know that the constraints imposed by the open-boundary conditions,
Eqs. (13) to (15), mean that every quantity on the right-hand side
of the compressibility relation is independent of the current $I$.
The open-system compressibility sum rule follows directly:

\item{$\bullet$} {\it in an open mesoscopic channel, the
electronic compressibility is perfectly independent of the
applied current and is given by Eq. (16).}

\noindent
This statement has profound implications for the proper
description of mesoscopic transport. Since it is a strict
identity, which originates in the {\it charge-conserving structure}
of an open channel together with its interfaces,
the generalized compressibility sum rule has the same
canonical importance as the fluctuation-dissipation theorem.

To be meaningful and credible, a model of mesoscopic transport
and noise must satisfy not only the fluctuation-dissipation theorem,
but the nonequilibrium compressibility sum rule as well.
We will not discuss here how the Landauer-like theories
of mesoscopic noise fare with regard to this principle;
interested readers can find our analysis in Reference [21].
Such theories do not respect the compressibility sum rule.
Therefore the basis for their conclusions about fluctuation
properties is questionable.

\vskip 12 truept
\centerline{\it c. General Consequences for Noise}
\vskip 12 truept

Equipped with the knowledge of Eq. (16), we can make some
powerful inferences about the structure of the fluctuations
in a quantum point contact, when it is driven away from
equilibrium. In the limit of low density in the channel,
the carriers are classical. Then $\Delta N \to N$ and

$$
\kappa \to \kappa_{\rm cl} = {1\over nk_{\rm B}T}.
\eqno(17a)
$$

\noindent
In the opposite limit, the electrons are highly degenerate.
Since $n = 2m^*v_{\rm F}/\pi\hbar \propto \sqrt{\mu - \varepsilon_i}$
for subband $i$, then

$$
\kappa \to \kappa_{\rm cl}
{\left( {k_{\rm B}T\over n} {\partial n\over \partial \mu} \right)}
= \kappa_{\rm cl} {k_{\rm B}T\over 2(\mu - \varepsilon_i)}
\ll  \kappa_{\rm cl}.
\eqno(17b)
$$

\noindent
Not surprisingly the Pauli exclusion principle strongly
inhibits the scale of the fluctuations as the compressibility
becomes negligible, in comparison with a classical system
at the same density.

This tells us that degeneracy must also have an enormous effect on
the scale of the noise exhibited by a QPC (or any mesoscopic
device, for that matter). At high density the noise must scale
as in Eq. (17b), {\it no matter how hard the channel is driven}.
That constraint is enforced by the compressibility sum rule.
We therefore expect that the expression for nonequilibrium
noise in the QPC must always carry a proportionality to the
ratio $\kappa/\kappa_{\rm cl}$. A noise formula that behaved
otherwise would immediately advertise its violation of the
compressibility sum rule [21].

\vskip 12 truept
\centerline{\it d. Nonequilibrium Noise of a QPC}
\vskip 12 truept

Our treatment of QPC transport and noise
begins with the Boltzmann-Landau-Silin transport
equation [7] for electrons in a one-dimensional channel.
Since a qualitative appreciation of the sum-rule behavior is
enough to interpret the results, we merely sketch the basic
makeup of the theory and point readers to the full technical
details in Refs. [11], [13], and [22].

In the leads, outside the active region $-L/2 < x < L/2$,
the electrons are permanently undisturbed.
Within the leads, their distribution
function $f_k(x,t)$ is displaced from equilibrium by the
current injected at the source boundary, and extracted at the
drain. the nonequilibrium function satisfies

$$
{\left[ {\partial\over \partial t} + v_k {\partial\over \partial x}
+ {eE\over \hbar} {\partial\over \partial k}  \right]} f_k(x,t)
= -{\cal W}_k[f(t)].
\eqno(18)
$$

\noindent
The driving field $E$ is due to the resistivity
dipole [15] between the source and drain interfaces,
set up in response to the perturbing external current; see Fig. 1.
The collision term ${\cal W}[f(t)]$ contains all of the scattering
effects at the microscopic level, be they single- or multi-particle.
In the spirit of Boltzmann, Landau, and Silin it is represented
as a functional of the single-particle distribution.

We parametrize ${\cal W}[f(t)]$ in terms of the inelastic and
elastic collision rates $1/\tau_{\rm in}$ and $1/\tau_{\rm el}$,
in such a way that Eq. (18) remains exactly conserving [22].
We then solve the equation for $f_k(x,t)$ and its fluctuation
counterpart

$$
\Delta f_k(x,t) = k_{\rm B}T {\delta f_k\over \delta \mu}(x,t)
\equiv \int^{L/2}_{-L/2} dx' \int {2dk'\over 2\pi}
{\delta f_k(x,t)\over \delta f^{\rm eq}_{k'}(x')}
\Delta f^{\rm eq}_{k'}(x').
\eqno(19)
$$

\noindent
Eq. (19) contains the variational Green function
${\delta f_k(x,t)/\delta f^{\rm eq}_{k'}(x')}$.
This calculable object has all of the information needed to
construct the current-current correlator that determines
the noise spectrum [11, 22].

\topinsert
\vskip -0.30truecm
\input psfig.sty
\centerline{\hskip5mm\psfig{figure=dgtf3.ps,height=8.0truecm}}
%
\vskip 0.25truecm 

\noindent
{\bf Figure 3.}
Left scale: excess thermal noise $S^{\rm xs}$
of a ballistic QPC at source-drain voltage $V = 9k_{\rm B}T/e$,
normalized to the Johnson-Nyquist level at ideal conductance,
and plotted as a function of chemical potential [22].
Right scale: the corresponding two-point conductance $G$.
At the subband crossing points of $G$, the excess noise peaks.
Noise is high at the crossing points, where
subband electrons are {\it classical}, and low at the plateaux
where subband {\it degeneracy} is strong.
Much more than the conductance, $S^{\rm xs}$
is sensitive to the scattering-time ratio
$\zeta = \tau_{{\rm in}}/\tau_{{\rm el}}$.
There is a dramatic lowering of the upper noise peak
(chain and dotted lines)
as the inelasticity in the second subband is made stronger.  
Dashed line: ideal excess noise, including shot noise,
predicted by the Landauer-B\"uttiker model [5]
and corresponding to our full line
($\zeta_1 = \zeta_2 = 1$). The estimated noise is much smaller.
\vskip 8truept
\endinsert
\hsize = 15.5truecm

We present in Fig. 3 the analytic results of our nonequilibrium
noise model, based on strictly conservative kinetics. We show
both the QPC conductance $G$, whose form is identical to that
given by Eq. (6) with (9a), and the excess noise of the channel.
This represents the excitation strength of the current fluctuations
over and above the Johnson-Nyquist equilibrium contribution
$S_{\rm JN} = 4Gk_{\rm B}T$. At current $I$ and source-drain
voltage $V = I/G$, the expression for excess noise at fixed $V$ is

$$
S^{\rm xs}(V,\mu)
= {\left( {2e^2V^2\over m^*L^2} \right)}
{\kappa\over \kappa_{\rm cl}} G
{\left( \tau^2_{\rm in}
+ 2{\tau_{\rm el}\tau^2_{\rm in}\over {\tau_{\rm el}+\tau_{\rm in}}}
- {\tau^2_{\rm el}\tau^2_{\rm in}\over (\tau_{\rm el}+\tau_{\rm in})^2 }
\right)}.
\eqno(20)
$$

\noindent
To interpret Fig. 3 we need first to note the second and
third factors on the right-hand side, $\kappa/\kappa_{\rm cl}$
and $G$. The excess noise is plotted
as a function of chemical potential; that is, the channel density
goes from being practically depleted to where both subbands in the
model are fully degenerate.

How do these two factors behave in the transition from
classical to strongly quantum regimes for the electrons?
As we saw above, the compressibility ratio reaches its maximum
of unity in the classical limit, and decreases monotonically
as the density rises, vanishing as
$k_{\rm B}T/2(\mu - \varepsilon_i)$ in the degenerate limit.
At the same time, the conductance {\it rises} monotonically
from its exponentially small value
$\sim e^{-(\varepsilon_i - \mu)/k_{\rm B}T}$
to its maximum Landauer value of $G_0$ per filled subband.
At each threshold $\mu \approx \varepsilon_i$, the two countervailing
trends combine to generate the strong peaks seen in the Figure. 

Next we consider the last factor in Eq. (20). It is a function
of the inelasticity ratio $\zeta = \tau_{\rm in}/\tau_{\rm el}$,
and a highly responsive one at that.
As $\zeta$ is decreased for the upper subband, the conductance
shows nonideal behavior just as in Fig. 2. However,
the corresponding suppression of the second noise peak is
dramatic. In practical terms this means the noise is an
{\it extremely sensitive predictive marker} of the inelastic
processes inside the QPC, much more so than the shot-noised
based prediction of the phenomenological models [5],
which reflects only the change in $G$.

Before applying our model to an experimental situation,
we go back to the question of whether changes in $G$ are truly
governed by inelastic scattering via $\zeta$ or whether,
as required by the Landauer theory, it depends on elastic
backscattering alone to modify $\tau_{\rm el}$
while dissipative effects never enter the picture.

Also in Fig. 3 we plot the {\it ideal}, shot-noise
based prediction for the excess noise prescribed by the
Landauer-B\"uttiker (LB) theory of QPC fluctuations [5].
At the relatively large source-drain voltage used to
calculate our results, the LB prediction is far smaller
than our ideal result (topmost set of peaks). Only for
enhanced inelasticity, $\zeta < 1$, does our fully conserving
kinetic calculation of $S^{\rm xs}$ begin to come down
to the LB curve; at small driving voltages the situation
is no different [22].
This graphic comparison is quite apart from the
fact that the LB model fails to respect the compressibility sum
rule [21], as shown by the absence of $\kappa$
anywhere in the Landauer-B\"uttiker noise formula [5].

\vskip 12 truept
\centerline{\it e. Comparison with Experiment}
\vskip 12 truept

In 1995, a significant set of QPC noise measurements was
published by Reznikov, Heiblum, Shtrikman, and Mahalu [12].
This experiment was distinguished by the fact that,
for the first time, such noise measurements were performed
for a range of fixed values of the current as well as for
fixed source-drain voltages, as is more customary
(and for which the outcome accords semi-quantitatively
with Fig. 3 above).

\topinsert
\vskip -0.30truecm
\input psfig.sty
\centerline{\hskip00mm\psfig{figure=dgtf4.ps,height=8.5truecm}}
%
\vskip 0.25truecm 

\noindent
{\bf Figure 4.}
Nonequilibrium current noise in a quantum point contact at 1.5K,
after the measurements of Reznikov {\it et al}., Ref. [12]
as a function of gate bias, at fixed source-drain current values.
The gate bias allows the electron density in the channel to
be swept from near-depletion to full degeneracy.
Dotted line: the most widely adopted theoretical noise model [5]
typically produces a strictly monotonic noise signal at
the first subband energy threshold.
That model totally fails to predict the very strong noise
peaks actually observed at threshold.
\vskip 8truept
\endinsert
\hsize = 15.5truecm

Figure 4 reproduces the fixed-current noise plots from Ref. [12].
On the basis of the LB formula, one would have anticipated
a {\it strictly monotonic} behavior in $S^{\rm xs}$,
complementary to the rising stepwise form of $G$ as the
gate voltage applied to the mesoscopic channel sweeps its
electron density upwards into the degenerate limit.
No monotonic falloff is seen in the experimental data.

Unequivocally, the Reznikov {\it et al}. experiment at
fixed values of QPC current tells us that the
Landauer-B\"uttiker noise model -- along with its many
emulators -- does not work. If it did, the dramatic peak
features of Fig. 4 would have been predicted.
Something essential is evidently missing from the
phenomenological treatments. It is the vital action of the
compressibility sum rule, whose noise signature we have
already prefigured and analyzed in Fig. 3 above.

Just such a family of constant-current noise maxima results from
our conserving kinetic theory for the excess noise [13].
We show our calculation of the noise peaks in Fig. 5.
They are in excellent accord with the structures
measured by Reznikov {\it et al.} around the subband threshold.$^5$
\footnote{}{$^5$ Neither the LB model, nor our presently
simplified one, reproduces the measured noise plateaux
in the extreme low-density limit, where the electrons are
classical and highly accelerated by the enormous applied field,
which by then goes as $V \sim I/G$ while $G \to 0$.
A proper account of the dynamics in this extreme situation
would have to cover the transition between the one-dimensional
physics in the QPC and that in a two- or even three-dimensional
scenario, as the energized electrons break out of
their confinement within the QPC channel.}

\topinsert
\vskip -0.30truecm
\input psfig.sty
\centerline{\hskip00mm\psfig{figure=dgtf5.ps,height=8.0truecm}}
%
\vskip 0.25truecm 

\noindent
{\bf Figure 5.}
Excess hot-electron noise at 1.5K in a QPC
at its first subband threshold, computed
with our strictly conserving Eq. (21),
as a function of chemical potential
(referred to the first threshold)
at fixed levels of source-drain current.
(See our Ref. [13]).
There is close quantitative affinity of
our peaks to the experimentally observed
first-threshold maxima in Fig. 4 of Ref. [12].
The dotted line at 100nA shows the Landauer-B\"uttiker
prediction [5] based on Eq. (6). This should be
compared with the chain line based on Eq. (21).
\vskip 8truept
\endinsert
\hsize = 15.5truecm

What generates the noise maxima in the case of Fig. 5?
The indispensable role of compressibility is as before:
the ratio $\kappa/\kappa_{\rm cl}$ moves smoothly from
unity at low density, to vanishing values at large $n$.
Interestingly, in the case of constant current, the competing
effect that leads to the threshold peaks is different to
that for constant source-drain voltage, where $G$ filled
that role. Here, it is the last factor in Eq. (20), which
as we saw in Fig. 3 is extremely sensitive to inelasticity.

In an experimental situation where the energy $eV = eI/G$,
available to drive the carriers, gets larger and larger
as the density goes down, we must expect inelastic phonon
emission to become dominant. Hence the inelastic relaxation
rate $1/\tau_{\rm in}(I, \mu)$ increases markedly, and
the inelasticity parameter $\zeta(I, \mu)$ drops fast.
Let us recast Eq. (20) at fixed $I$:

$$
S^{\rm xs}(I,\mu)
= {\left( {2e^2I^2\tau_{\rm el}^2\over m^*L^2} \right)}
{\kappa\over \kappa_{\rm cl}}
{\zeta(I,\mu)^2\over G(I, \mu)}
{\left( 1 + {2\over {1 + \zeta(I,\mu)}}
- {1\over (1 + \zeta(I,\mu))^2 } \right)}.
\eqno(21)
$$

\noindent
Omitting details of our calculation (see Ref. [13]),
we discuss this form qualitatively. Attention falls on
the behavior of those quantities that depend strongly
on the field, and hence on $I$. The elastic relaxation
rate $1/\tau_{\rm el}$ is relatively immune to field effects;
it remains as in Eq. (7a).

\item{1.} {\it ``Pinchoff'' limit}. (This is when a large
negative bias on the control gate depletes the QPC channel.)
Since $G \sim n\zeta$ while $\kappa/\kappa_{\rm cl} = 1$,
we have that

$$
S^{\rm xs}(I,\mu) \sim I^2 {\zeta(I,\mu)/n}.
$$

\item{}
When the field-induced inelasticity is very strong,
$\zeta$ falls faster than the density. The net effect
is to cause the excess noise to go down.

\item{2.} {\it Degenerate limit}.
Well above the threshold, the conductance is near-ideal
as the energy $eV$ is less
than the Fermi energy $\mu - \varepsilon_1$. Then $\zeta \approx 1$
and we recover the same asymptotic behavior as before, where
$S^{\rm xs}$ is dominated by $\kappa \sim 1/(\mu - \varepsilon_1)$,
and therefore falls off at high density.

\noindent
There must be a crossover between these two extremes, and we see
it precisely in the region of transition between classical and
quantum degenerate regimes. That transition occurs at threshold.

\topinsert
\vskip -0.30truecm
\input psfig.sty
\centerline{\hskip00mm\psfig{figure=dgtf6.ps,height=7.0truecm}}
%
\vskip 0.25truecm 

\noindent
{\bf Figure 6.}
Hot-electron excess noise calculated for a quantum
point contact at 1.5K,
for fixed values of source-drain driving voltage $V$
(going upwards, the values are 0.5, 1, 1.5, 2, and 3meV).
Normalization of the noise is to the thermal value
$2eI_{\rm th} \equiv 2G_0k_{\rm B}T$.
The peak heights rise monotonically and just less
than linearly with $V$.
Nonideality from the inelastic scattering is strong enough
to suppress the thermal peaks (refer to Fig. 3), making
them quasilinear in the voltage. The maxima
agree well with experimental results (see Fig. 2, Ref. [12]).
Our fully conserving, nonequilibrium kinetic computation
shows that {\it quasilinear dispersion of the noise maxima
with $V$ is not unique to shot noise}.
\vskip 8truept
\endinsert
\hsize = 15.5truecm

\topinsert
\vskip -0.30truecm
\input psfig.sty
\centerline{\hskip00mm\psfig{figure=dgtf7.ps,height=7.0truecm}}

\noindent
{\bf Figure 7.}
The total QPC conductance $G$ corresponding to Fig. 6.
Our results are comparable to the measurements of
Reznikov {\it et al.} [12] (see their Fig. 2).
Our choice of subband spacing, $7k_{\rm B}T$,
approximately equals the shoulder width of the steps
as noted for the experimental plots of $G$s.
The step-like size of $G$ decreases monotonically with the
applied voltage (as in Fig. 6 but reading downwards from 0.5 to 3meV).
There is progressively greater loss of ideality
as $eV$ exceeds the subband gap energy
$\varepsilon_2 - \varepsilon_1 = 0.9$meV,
and inelastic phonon emission sets in.
\vskip 8truept
\endinsert
\hsize = 15.5truecm

To end our discussion, we present in Figs. 6 and 7 the results
of our model for constant $V$. Fig. 6 shows the noise, while
Fig. 7 shows the conductance of a QPC. Our calculation of
$S^{\rm xs}(V, \mu)$ fits the fixed-voltage data of Ref. [12]
quite well, including the quasi-linear dispersion of the
threshold peak heights, as functions of $V$. While such
dispersion is popularly considered as unique to shot noise,
we emphasize
that {\it there is no shot noise in Eqs. (20) and (21)}.
Our microscopically derived excess noise formula describes
so-called hot-electron noise which is an entirely different
quantity, thermodynamically, from shot noise [11].
Had we been describing shot noise, we would not have seen
the governing influence of the compressibility on the results.
The evidence for the role of the compressibility sum rule
is in Fig. 4, the outcome of a real experiment. Shot noise
cannot account for it; hot-electron noise does, fully.

\vskip 16 truept
\centerline{\bf 5. FUTURE APPLICATIONS}
\vskip 12 truept

A potentially rich field of investigation lies within the
structure of the inelasticity $\zeta$ and its mode of
interaction with the compressibility sum rule, leading
to the striking form of the excess noise spectrum.
Inelasticity is not only a function of experimental control
parameters such as current or applied voltage.
Far more important is its dependence on both {\it materials}
and {\it device geometry}.

As an example of what can be done with the tools provided
by Eq. (20) and our supporting kinetic machinery, consider
the structural differences between a QPC channel embedded
in a heterojunction substrate, as in the Reznikov {\it et al}.
experiment [12], and a suspended carbon nanotube. Both are
essentially one-dimensionally confined channels. However,
in the former case, inelastic phonon emission couples the
carriers to a bath of three-dimensional lattice excitations.
In the latter case both the electrons {\it and} the phonons
are one-dimensional.

We therefore predict qualitatively different forms of
hot-electron noise behavior from  experiments done on
suspended nanotubes. Such noise measurements have already been
performed [23], though not at fixed current as far as we
are aware. Because the behavior of $\zeta$ will be quite
different, it should be interesting to compare $S^{\rm xs}$
measured for suspended tubes, with other tubes intimately
contacted to, or even embedded in, a surrounding matrix.

Finally we note that the electronic compressibility
$\kappa$ is a strong function of the electron-electron
correlations in a metallic electron system [6].
Such correlations are enhanced at low temperatures
and low densities. The information on electron correlations,
conveyed by the compressibility, should then be
available through studies of nonequilibrium noise in structures
where those correlations become significant. Comparisons
could be made between, say, sound-velocity data and noise data.

\vskip 16 truept
\centerline{\bf 6. SUMMARY}
\vskip 12 truept

Our review has covered much territory.
We began with some background to contemporary developments in
mesoscopic transport, and raised a number of delicate points
which modern phenomenological approaches do not address.
We posed two questions: do such philosophies of transport
(and noise) still fall short in their
theoretical treatment of electron motion at short scales? If so,
is there still something novel to learn from the body of
knowledge firmly in place well before Landauer?

We then focused on the microscopic derivation of the
Landauer conductance formula, paying close attention to
the need to integrate charge conservation consistently
into the dynamical description of an {\it open} mesoscopic
channel. We found that those boundary conditions that are
uniquely consistent with the fluctuation-dissipation theorem
(to which the Kubo-Greenwood formula is equivalent)
require active sources and sinks to carry the external current
into and out of the driven device [10, 19].
{\it These are not the boundary conditions invoked
by quasi-diffusive phenomenologies of transport}.

Nevertheless, the Landauer conductance formula emerged very
naturally from the microscopic Kubo-Greenwood theory applied
to a quantum point contact. We identified the role of
{\it active dissipation} of electrical energy, effected by the
inevitable presence of inelastic scattering in the leads,
as a key ingredient in the microscopic derivation. This resolves
the strange cognitive gap in quasi-diffusive models, whereby a purely
coherent -- nondissipative -- description of conductance must
unaccountably ``produce'' dissipative energy loss not on the
basis of physics, but solely to save the appearance of
the fluctuation-dissipation relation within Landauer's theory.

At this point we were able to answer the two initial questions
in the specific context of transport. First, Landauer-like approaches
to mesoscopic transport, albeit necessary and highly successful
correctives to inappropriate bulk averaging at small scales, remain
microscopically incomplete. Second, their completion comes {\it only}
through the standard microscopic theory of electrically open systems,
unaided by any superfluous phenomenology. The new ingredient here
is the explicit appearance of incoherent, inelastic scattering.

Noise was the next phenomenon examined. The very same
canonical boundary conditions, responsible for the microscopically
consistent derivation of the Landauer formula, turned out to constrain
the electron-hole pair fluctuations arising in the conductive channel.
The constraint is in the form of the {\it compressibility sum rule},
itself a consequence of global charge conservation. The constraint
is so strong that it compels the noise of a degenerate channel to
scale with the ratio of the thermal bath energy to the Fermi energy
{\it no matter how hard the system is driven by an applied current.}

A novel outcome of the compressibility sum rule emerged from
our detailed kinetic-theoretical treatment of nonequilibrium noise
in a QPC. The interplay of compressibility and the scattering dynamics
within the channel led to a characteristic series of noise peaks,
which we were able to match finely to the measurements
of Reznikov {\it et al}. [12]. With the help of the compressibility
sum rule we have successfully explained [13] the remarkable noise
features observed at fixed levels of the driving current.
Such structures have no explanation in terms of more
phenomenological, quasi-diffusive noise models [5].

We have revealed the marked, and previously unsuspected, influence
of the conserving sum rules in the phenomena surrounding mesoscopic
electron motion. This is particularly true of the compressibility sum
rule which, beyond QPCs, may be expected to play an equally pre-eminent
role over the whole stretch of mesoscopic transport and noise problems,
and at scales yet smaller [24]. A range of novel experimental effects
can now be theoretically discussed and sought out in appropriately
planned observations. Noise in carbon nanotubes is only one of a
potentially wide set of options to explore.

Questions about the microscopic standing of modern phenomenologies
are not unprecedented. They have been repeatedly raised within the
mesoscopics community (and repeatedly avoided) from the
earliest times that the coherent-cum-diffusive picture was taken
up in a major way. Our own considerations strike out in new and
different directions, but they are also consonant with a whole family
of other critiques. Aside from the impressive formal contributions
by Sols [10] and the IMEC group [19, 25], readers may wish
to refer to elegant papers by Fenton [26] and Kamenev and Kohn [27],
and to the overview of established mesoscopic theories contained
in the recent review by Agra\"{\i}t {\it et al.} [24].

\vskip 16 truept
\centerline{\bf REFERENCES}
\vskip 12 truept

\item {[1]}
D. K. Ferry and S. M. Goodnick, {\it Transport in Nanostructures}
(Cambridge University Press, Cambridge, 1997).

\item{[2]}
S. Datta, {\it Electronic Transport in Mesoscopic Systems}
(Cambridge University Press, Cambridge, 1997).

\item{[3]}
J. H. Davis, {\it The Physics of Low Dimensional Semiconductors: an
Introduction}, (Cambridge University Press, Cambridge, 1998).

\item{[4]}
Y. Imry, {\it Introduction to Mesoscopic Physics} second edition
(Oxford University Press, Oxford, 2002).

\item{[5]}
Y. M. Blanter and M. B\"uttiker,
{\it  Phys. Rep.} {\bf 336}, 1 (2000).

\item{[6]}
D. Pines and P. Nozi\`eres, {\it  The Theory of Quantum Liquids},
(Benjamin, New York, 1966).

\item{[7]}
A. A. Abrikosov, {\it  Fundamentals of the Theory of Metals}
(North-Holland, Amsterdam, 1988).

\item{[8]}
R. Kubo, M. Toda, and N. Hashitsume,
{\it Statistical Physics II: Nonequilibrium Statistical Mechanics},
second edition (Springer, Berlin, 1991).

\item{[9]}
J. M. Ziman {\it Models of Disorder}
(Cambridge University Press, Cambridge, 1979), Ch. 10.

\item{[10]}
F. Sols, {\it  Phys. Rev. Lett.} {\bf 67}, 2874 (1991).

\item{[11]}
F. Green and M. P. Das,
{\it  J. Phys.: Condens. Matter} {\bf 12}, 5251 (2000).

\item{[12]}
M. Reznikov, M. Heiblum, M. Shtrikman, and D. Mahalu,
{\it Phys. Rev. Lett.} {\bf 75}, 3340 (1995).

\item{[13]}
F. Green, J. S. Thakur, and M. P. Das,
{\it Phys. Rev. Lett.} {\bf 92}, 156804 (2004).

\item{[14]}
R. Landauer, {\it IBM J. Res. Dev.} {\bf 1}, 223 (1957);
{\it Phil. Mag.}{\bf 21}, 863 (1970).

\item {[15]}
Y. Imry and R. Landauer, {\it Rev. Mod. Phys.} {\bf 71}, S306 (1999).

\item{[16]}
R. Landauer, in {\it Coulomb and Interference
Effects in Small Electronic Structures},
ed. by D. C. Glattli, M. Sanquer and J. Tran Than Van
(Editions Fronti\`eres, Gif-sur-Yvette, 1994) p. 1.

\item{[17]} 
S. Doniach and E. H. Sondheimer,
{\it Green Functions for Solid State Physicists}
(W. A. Benjamin, Reading, MA, 1974).

\item{[18]}
N. G. Van Kampen, {\it Stochastic Processes in Physics and Chemistry}
(North-Holland, Amsterdam, 1992)

\eject
\item{[19]}
W. Magnus and W. Schoenmaker, {\it Quantum Transport in
Sub-micron Devices: A Theoretical Introduction}
(Springer, Berlin ,2002).

\item{[20]}
M. P. Das and F. Green, {\it  J. Phys.: Condens. Matter}
{\bf 15} L687 (2003).

\item{[21]}
M. P. Das, J. S. Thakur, and F. Green, {\it Int. J. Mod. Phys. B},
to appear; see also arXiv preprint cond-mat/0401134.

\item{[22]}
F. Green and M. P. Das, {\it  Fluct. Noise Lett.} {\bf 1}, 21 (2001).

\item{[23]}
P.-E. Roche, M. Kociak, S. Gu\'eron,
A. Kasumov, B. Reulet, and H. Bouchiat,
{\it Eur. Phys. J. B} {\bf 28}, 217 (2002).

\item{[24]}
N. Agra\"{\i}t, A. Levy Yeyati, and J. M. van Ruitenbeek,
{\it Phys. Rep.} {\bf 377}, 81 (2003); see Sec. III.D.5.

\item{[25]}
B. Sor\'ee, Ph. D. thesis, Leuven, 2003;
B. Sor\'ee, W. Magnus, and W. Schoenmaker,
{\it Phys. Lett. A} {\bf 310}, 322 (2003).

\item{[26]}
E. W. Fenton, {\it Phys. Rev. B} {\bf 46}, 3754 (1992);
{\it Superlattices and Microstruct.} {\bf 16}, 87 (1994).

\item{[27]}
A. Kamenev and W. Kohn,
{\it Phys. Rev. B} {\bf 63}, 155304 (2001).


\end

\end{document}